\documentclass[a4paper,fleqn,usenatbib]{mnras}
\usepackage{pdflscape}

\usepackage[percent]{overpic}
\usepackage{environ}
\usepackage{multirow,array,booktabs}
\usepackage{lmodern}
\usepackage[export]{adjustbox}

\newdimen\fontdim
\newdimen\upperfontdim
\newdimen\lowerfontdim
\newif\ifmoreiterations
\makeatletter
\NewEnviron{fitbox}[2]{
  \def\buildbox{%
    \setbox0\vbox{\hbox{\minipage{#1}%
      \fontsize{\fontdim}{1.2\fontdim}%
      \selectfont%
      \stuff%
    \endminipage}}%
    \dimen@\ht0
    \advance\dimen@\dp0
  }
  \def\stuff{\BODY}
  \buildbox
  \ifdim\dimen@>#2
    \loop
      \fontdim.5\fontdim 
      \buildbox
    \ifdim\dimen@>#2 \repeat
    \lowerfontdim\fontdim
    \upperfontdim2\fontdim
    \fontdim1.5\fontdim
  \else
    \loop
      \fontdim2\fontdim 
      \buildbox
    \ifdim\dimen@<#2 \repeat
    \upperfontdim\fontdim
    \lowerfontdim.5\fontdim
    \fontdim.75\fontdim
  \fi
  \loop
    \buildbox
    \ifdim\dimen@>#2
      \moreiterationstrue
      \upperfontdim\fontdim
      \advance\fontdim\lowerfontdim
      \fontdim.5\fontdim
    \else
      \advance\dimen@-#2
      \ifdim\dimen@<10pt
        \lowerfontdim\fontdim
        \advance\fontdim\upperfontdim
        \fontdim.5\fontdim
        \dimen@\upperfontdim
        \advance\dimen@-\lowerfontdim
        \ifdim\dimen@<.2pt
          \moreiterationsfalse
        \else
          \moreiterationstrue
        \fi
      \else
        \moreiterationsfalse
      \fi
    \fi
  \ifmoreiterations \repeat
  \box0
}
\makeatother

\usepackage[T1]{fontenc}
\usepackage{ae,aecompl}

\usepackage{graphicx}	
\usepackage{amsmath}	
\usepackage{amssymb}	
\usepackage{enumitem}
\usepackage [authoryear]{natbib}
\usepackage[outdir=./]{epstopdf}
\usepackage{color}

\title[Transition in galaxy star formation efficiency]{VALES: IV. 
Exploring the transition of star formation efficiencies between normal and starburst
galaxies using APEX/SEPIA Band-5 and ALMA at low redshift}

\author[C. Cheng et al.]{C. Cheng$^{1,2}$, 
E. Ibar$^{1}$, 
T. M. Hughes$^{1,2,3,4}$, 
V. Villanueva$^{1}$, 
R. Leiton$^{1}$,  
G. Orellana$^{1}$, 
\newauthor
A. Munoz-Arancibia$^{1}$, 
N. Lu$^{2}$,
C. K. Xu$^{2}$,
C. N. A. Willmer$^{5}$,
J. Huang$^{2}$,
T. Cao$^{2}$,
\newauthor
C. Yang$^{6,7,8}$,
Y. Q. Xue$^{3,4}$
and K. Torstensson$^{6}$
\\
$^{1}$Instituto de F\'isica y Astronom\'ia, Universidad de Valpara\'iso, Avda. Gran Breta\~na 1111, Valpara\'iso, Chile. Email: cheng.cheng@uv.cl\\
$^{2}$
Chinese Academy of Sciences South America Center for Astronomy, China-Chile Joint Center for Astronomy, \\
	National Astronomical Observatories, Chinese Academy of Sciences, Beijing 100012, China\\
$^{3}$CAS Key Laboratory for Research in Galaxies and Cosmology, Department of Astronomy, 
University of Science and Technology of China, Hefei 230026, China\\
$^{4}$School of Astronomy and Space Science, University of Science and Technology of China, Hefei 230026, China\\
$^{5}$Steward Observatory, University of Arizona, 933 N. Cherry Avenue, Tucson, AZ 85721, USA\\
$^{6}$European Southern Observatory, Alonso de Cordova 3107, Casilla 19001, Santiago 19, Chile\\
$^{7}$Purple Mountain Observatory/Key Lab of Radio Astronomy, Chinese Academy of Sciences, Nanjing 210008, China\\
$^{8}$Institute d'Astrophysique Spatiale, CNRS, Univ. Paris-Sud, Universit\'e Paris-Saclay, B\^at. 121, 91405 Orsay Cedex, France
}

\date{Accepted XXX. Received YYY; in original form ZZZ}

\pubyear{2017}

\begin{document}
\label{firstpage}
\pagerange{\pageref{firstpage}--\pageref{lastpage}}
\maketitle

\begin{abstract}
	In this work we present new APEX/SEPIA Band-5 observations targeting the CO ($J=2\text{--}1$) 
	emission line of 24 Herschel-detected galaxies at $z=0.1-0.2$.
	Combining this sample {with} our recent new Valpara\'iso ALMA Line Emission Survey (VALES), 
	we investigate the star formation efficiencies (SFEs = SFR/$M_{\rm H_{2}}$) of galaxies 
	at low redshift. We find the SFE of our sample bridges the gap between normal 
	star-forming galaxies and Ultra-Luminous Infrared Galaxies (ULIRGs), which are thought 
	to be triggered by different star formation modes. {Considering the $\rm SFE'$ as the SFR and
	the $L'_{\rm CO}$ ratio, our data show a continuous and smooth increment as a function of infrared luminosity 
	(or star formation rate) with a scatter about 0.5 dex, instead of a steep jump with a bimodal behaviour}.
	{This result is due to the use of a sample with a much larger range
	of sSFR/sSFR$_{\rm ms}$ using LIRGs, with luminosities covering the range between normal
	and ULIRGs.} We conclude that the main parameters controlling the scatter {of the} 
	SFE {in star-forming galaxies} are the {systematic uncertainty of the} 
	$\alpha_{\rm CO}$ conversion factor, the gas fraction and physical size.
\end{abstract}




\section{Introduction}
The star formation efficiency (SFE) of a galaxy, defined as the ratio between the star formation rate (SFR) 
and the amount of gas reservoir, is a crucial parameter to characterise its star formation activity and future evolution. 
Previous studies have shown that the SFE of normal star-forming galaxies (SFGs) are systematically 
lower than that of starburst galaxies, suggesting different mechanisms could be triggering the star 
formation (Genzel et al. 2010; Daddi et al. 2010; Carilli and Walter 2013). 
This lead to a proposal of a bimodal SFE for the two types of star forming galaxies.
The first is associated with the 
``main sequence'' (Brinchmann et al. 2004) formed by normal SFGs, where the star formation is triggered within
disk-like structures, and their specific SFR (sSFR = SFR / $M_{*}$) slowly decreases with the 
increasing stellar mass (Schawinski et al. 2014). The second is associated with starburst galaxies that have typical sSFR
above the main sequence, and include local Ultra-Luminous Infrared Galaxies (ULIRGs, Solomon et al. 1997; Lonsdale et al. 2006). 
Nevertheless, it is unclear whether the SFE bimodality is caused by a higher gas-to-star conversion
rate or simply by a higher fraction of dense molecular gas capable of initiating star formation.
Galaxies with SFR in between the ULIRGs and the normal SFGs, e.g., Luminous Infrared Galaxies 
(LIRGs, $10^{11}$$<$$L_{\rm IR}/L_\odot$$<$$10^{12}$), are a critical population to understand the SFE bimodality.

In studying the SFE in LIRGs, one obstacle is how to measure the molecular hydrogen mass ($M_{\rm H_2}$).
Historically, CO has become the most popular tracer of interstellar cold molecular gas. 
One common method to translate the CO luminosity into the $M_{\rm H_2}$ is the adoption of a 
simple conversion factor $\alpha_{\rm CO} = M_{\rm H_{2}}/L'{\rm _{CO}}$. 
However, it is challenging to accurately define the value of $\alpha_{\rm CO}$ (Carilli and Walter 2013; Bolatto et al. 2013, and references therein). 
Empirically, the molecular gas clouds in the Milky Way and nearby galaxies show an $\alpha_{\rm CO}$ of 
$4.6\, M_{\odot}\,(\rm K\, km\, s^{-1} \,pc^{2})^{-1}$, which includes the helium correction (e.g., Solomon \& Vanden Bout 2005). 
If the molecular clouds in SFGs have similar properties such as metallicity, 
dynamical state, gas density, then the $\alpha_{\rm CO}$ of SFG shall 
resemble {that seen in local} molecular clouds (Solomon \& Vanden Bout 2005). On the other hand, 
simulations show that galaxy mergers change the cloud and inter-cloud properties such as the rising of 
the velocity dispersion and kinetic temperature, which increases the CO intensity (Narayanan et al. 2011). 
Thus the $\alpha_{\rm CO}$ of {interacting} galaxies should drop by a factor of 2-10 (Narayanan et al. 2012a). 
Detailed modelling and observational results suggest 
$\alpha_{\rm CO} = 0.8\, M_{\odot}\,(\rm K\, km\, s^{-1} \,pc^{2})^{-1}$ (Downes et al. 1998) as a consistency 
value to derive $M_{\rm H_2}$ in ULIRGs. 

Papadopoulos et al. (2012a, 2012b) found that the $\alpha_{\rm CO}$ in LIRGs can change significantly.
Previous studies show that the value of $\alpha_{\rm CO}$ in LIRGs can be similar to that in 
ULIRGs (Solomon et al. 1997; Yao et al. 2003), where $\alpha_{\rm CO}$ is lower than the Galactic value 
(e.g., $\alpha_{\rm CO} = 0.5\, M_{\odot}\,(\rm K\, km\, s^{-1} \,pc^{2})^{-1}$ in VV 114; 
$\alpha_{\rm CO} = 1.5\, M_{\odot}\,(\rm K\, km\, s^{-1} \,pc^{2})^{-1}$ in NGC 6240: Sliwa et al. 2013; Tunnard et al. 2015).
However, Papadopoulos et al. (2012a, 2012b) also find that the $\alpha_{\rm CO}$ in (U)LIRGs 
can have values close to {those} found in the Galaxy (Costagliola et al. 2013; Sandstrom et al. 2013). 
The kinetic gas components in LIRGs can mix different star formation activities: 
gas in LIRGs may show compact (Xu et al. 2014) as well as ring structures (Xu et al. 2015). 
This strongly indicates that the $\alpha_{\rm CO}$ may not have the same value for all LIRGs.

To explore the changing mechanism of the SFE between the ``main sequence'' and the starburst regimes, 
we observe a sample of LIRGs at 0.1$<$$z$$<$0.2 with the Swedish-ESO PI receiver for APEX (SEPIA, {Billade et al. 2012})
targeting the CO ($J=2\text{--}1$) emission line. We assume a cosmological model with $H_0 = 70\rm\, km/s/Mpc$, $\Omega_{\rm m}$ = 0.3, 
and $\Omega_{\Lambda}$ = 0.7.

\section{Galaxy Samples and Data}
\subsection{Sample selection}
\begin{figure}
\centering
\includegraphics[width=0.5\textwidth]{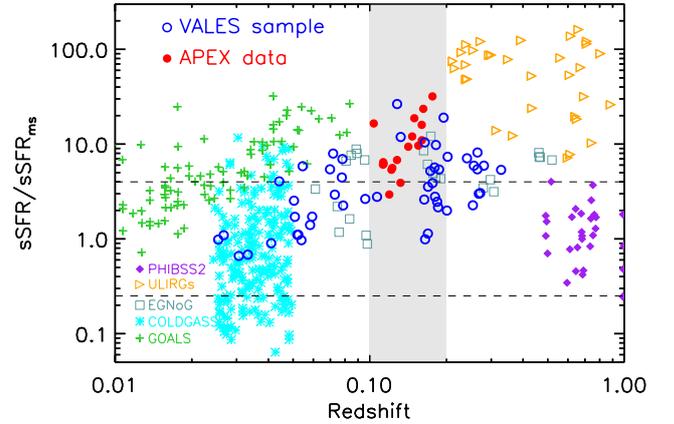}
\caption{The specific star formation rate (sSFR = SFR/$M_{*}$), normalised to that
estimated for the `main sequence' (MS, Elbaz et al. 2011; Whitaker et al. 2012) as a function of 
redshift for different samples of galaxies detected in CO. {We use the parameterisation of the MS as 
$\log[{\rm sSFR(MS,}z,{M_*)}] = -1.12 + 1.14z - 0.19z^2 - (0.3 + 0.13z) \times (\log M_* - 10.5)\, [\rm Gyr^{-1}]$ 
 (Genzel et al. 2015), where the dashed lines represent +/- 0.6 dex offsets from this relation for star-forming galaxies.}
Our VALES data are presented {as} blue open circles for ALMA and red solid circles for APEX observations. 
The green crosses are the IRS bright local galaxies  (GOALS: Armus et al. 2009);
the cyan asterisks are the nearby MS galaxies (COLDGASS: Saintonge et al. 2011); 
the light blue open squares are $z = 0.05-0.5$ normal galaxies (EGNoG: Bauermeister et al. 2013); the orange open triangles are the ULIRGs at intermediate redshifts (Combes et al. 2011; Combes et al. 2013);
the purple diamonds are the MS galaxies at 0.5$<$$z$$<$1  (PHIBSS: Combes et al. 2016).
More details about these archive data can be found in the Fig. 1 of Genzel et al. (2015).
The grey shaded region in this figure shows our target selection within 0.1$<$$z$$<$0.2,
which includes the CO ($J=1\text{--}0$) ALMA Band-3 detected galaxies and the CO ($J=2\text{--}1$) APEX/SEPIA Band-5 detected galaxies.}\label{sSFRms}
\end{figure}

The {\it Herschel} Astrophysical Terahertz Large Area Survey ($H$-ATLAS Eales et al. 2010)
covered 600 $\rm deg^2$ of the extragalactic sky with the PACS and SPIRE cameras
in the 100, 160, 250, 350 and 500 $\mu m$ bands. We selected targets from the 
{equatorial} $H$-ATLAS fields {covered by} the Galaxy And Mass Assembly
survey {which} has a rich multi-wavelength broadband coverage (Driver et al. 2016), including NUV and
FUV bands from the GALEX imaging, the optical images and spectroscopy from SDSS or GAMA, 
near-IR (NIR) imaging from the VISTA project, 
mid-IR (MIR) imaging from WISE and the {\it Herschel} far-IR (FIR) photometry. 

Making use of the public $H$-ATLAS DR1\footnote{\url{http://www.h-atlas.org/public-data/download}} 
catalog, we select sources with reliable optical counterpart, 
spectroscop{ic} redshift at 0.1$<$$z$$<$0.2, and located at the top of the sSFR distribution
 ($M_{*}$ from GAMA and SFR from $L_{\rm IR}$). These criteria allow weeding {out} the most 
intensely starbursting objects from the sample. Using these targets, we explore their molecular
gas content via their CO emission. These observations complement our recent
Valpara\'iso ALMA Line Emission Survey (VALES, Villanueva et al. 2017; Hughes et al. 2017a; 2017b),
which is the largest CO-detected galaxy sample at $z$$\sim$0.15. Fig. {1} shows the sSFR/sSFR$_{\rm ms}$
of the current CO-detected galaxy sample with different redshift. 
{There are several parametrisations of the main sequence galaxies 
(e.g., Speagle et al. 2014; Schreiber et al. 2015).} As the follow up work of our VALES
project, we adopt the sSFR = sSFR($M_{*}, z$) given by Genzel et al. (2015) as the VALES I paper (Villanueva et al. 2017).

\begin{figure*}
\centering
\includegraphics[width=0.95\textwidth]{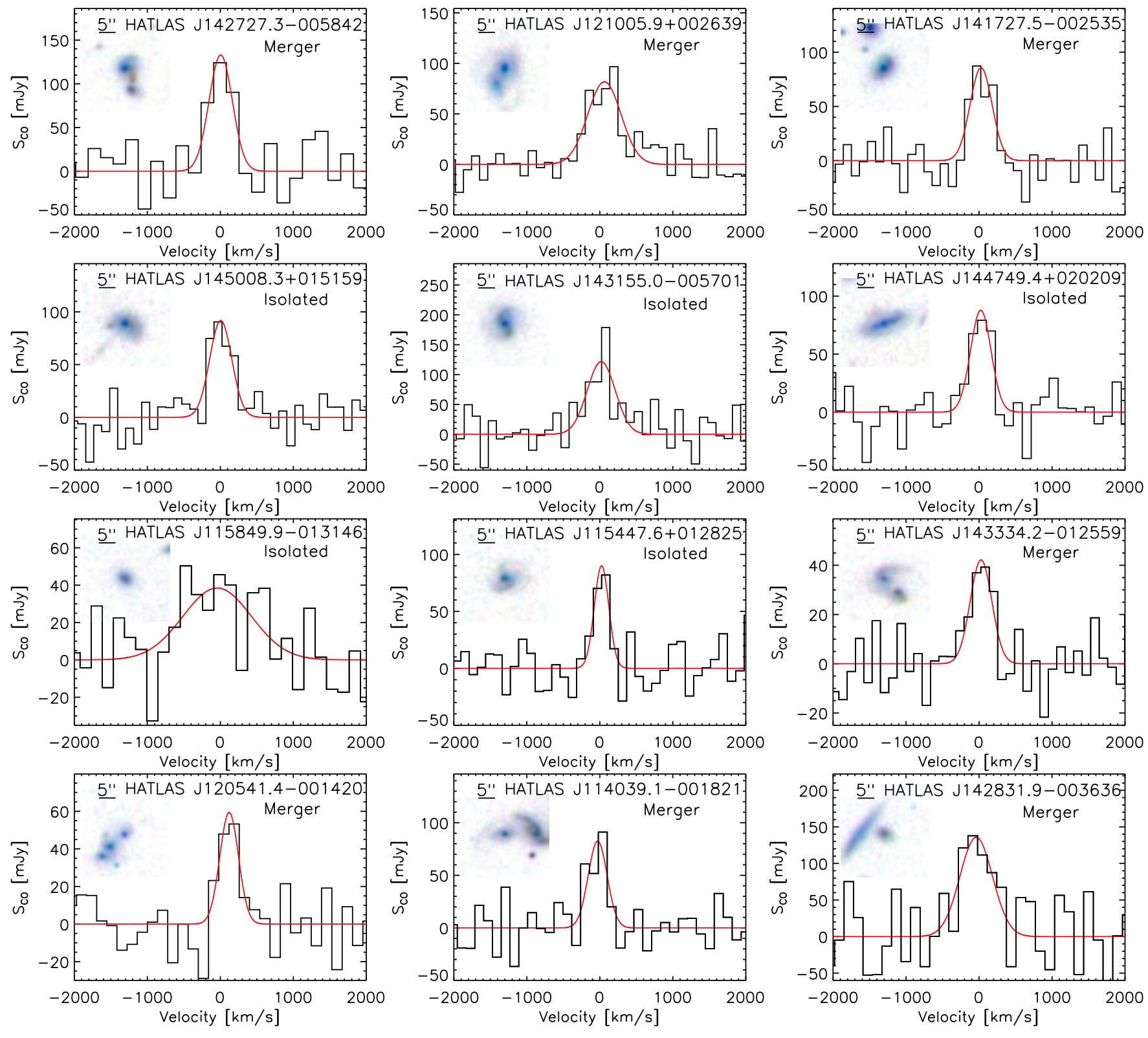}
\caption{The observed CO ($J=2\text{--}1$) spectra for APEX/SEPIA Band-5 
observed galaxies centred on spectroscopic redshifts taken from GAMA. The postage image in the 
upper left of each spectrum comes from the SDSS colour image with a scale of 25"$\times$25". 
The scale bar in the optical image denotes a length of 5''. 
The top 12 galaxies have CO detection with 5 $\sigma$.
We fit the CO emission by a single 
Gaussian profile and show the result as a red line. Typical FWHM is 300 km/s. 
HATLAS J115849.9-013146 is an isolated galaxy with fitted FWHM {of} about 1000 km/s, 
much larger than {that of} other galaxies. This galaxy is isolated and located in a group of three galaxies at 
$z = 0.147$ within 65 kpc ($\simeq 25.3''$). {The} Petrosian radii of these three galaxies are: 2.99'', 2.63'' and 2.89'', corresponding to a physical Petrosian diameter about 15 kpc. 
}\label{APEX}
\end{figure*}

\begin{figure*}
\centering
\includegraphics[width=0.95\textwidth]{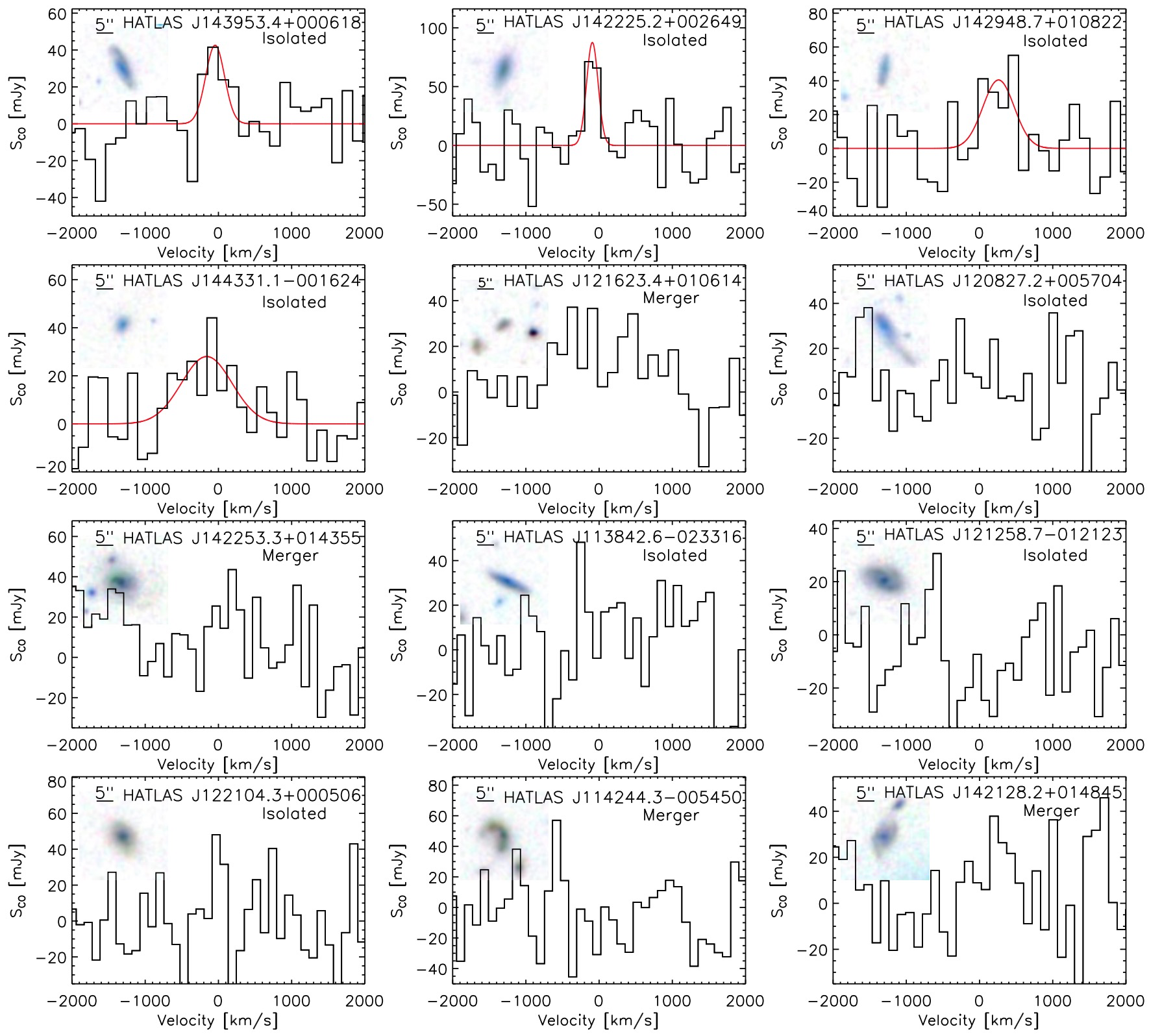}
{\textbf{Figure~{2}.} {\bf continued}: the first 4 targets have lower S/N between 3 and 5.
We also fit the CO emission by a single Gaussian profile and show the result as a red line.
The final 8 galaxies have no clear CO emission.}
\end{figure*}

\begin{figure*}
\centering
\includegraphics[width=0.9\textwidth]{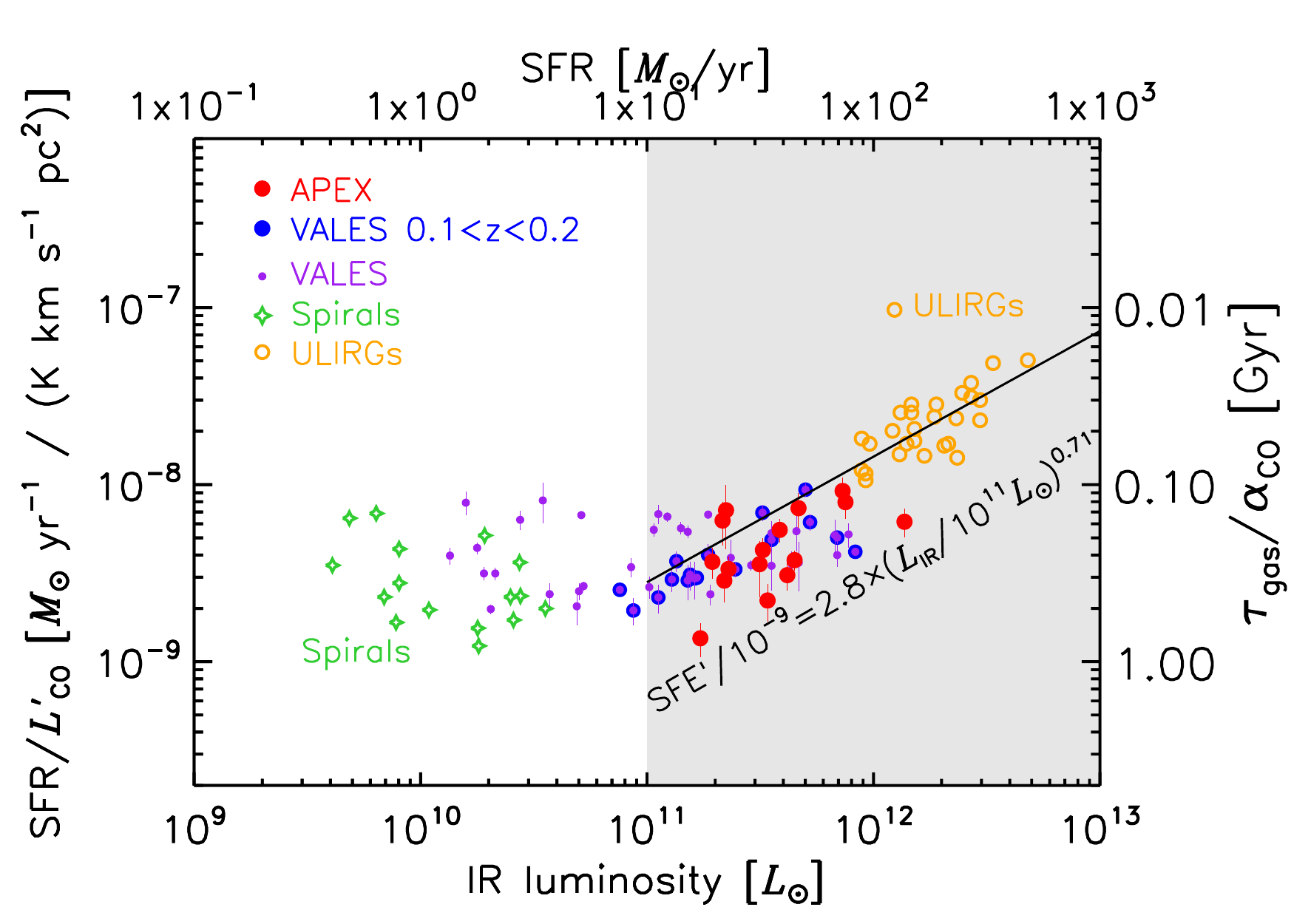}
\caption{
The ${\rm SFE}'$($={\rm SFR}/L'_{\rm CO}$) vs $L_{\rm IR}$ of our CO detected sample 
(blue and red dots); the VALES sample (purple/blue/red dots); local spirals
 (green stars; Leroy et al. 2009) and the local ULIRGs (orange cirlcles; Solomon et al. 1997). 
The top label shows the SFR scale.
The $L_{\rm IR}$ is derived from the {\it Herschel} data fitting as described in Villanueva et al. (2017). 
The right Y-axis is the gas depletion time-scale normalised by $\alpha_{\rm CO}$. The $\rm SFE'$ is nearly 
constant with 0.5 dex scatter at the IR luminosities between $2\times10^{9}L_{\odot}$ and $\times10^{11}L_{\odot}$. 
When galaxies are far-IR bright ($L_{\rm IR} > 10^{11}L_{\odot}$), 
the SFE tend to increase (see Equation {2}).}\label{SFE1}
\end{figure*}

\subsection{APEX/SEPIA Band-5 observations}
SEPIA Band-5 is a spectrograph that covers the frequency range 159-211 GHz, recently
mounted at {the Atacama Pathfinder Experiment (APEX)}. 
The water {absorption} line at 183 GHz is the main {feature} 
of the atmosphere at these frequencies. SEPIA Band-5 is the only instrument that 
is able to detect CO ($J=2\text{--}1$) at redshifts {from} 0.1 to 0.2 {in} 
the Southern hemisphere. We were awarded 21hrs of APEX/SEPIA Band-5 observations {to target} the CO ($J=2\text{--}1$) 
emission of 24 starburst galaxies (APEX programs 097.F-9724(A); 098.F-9712(B), PI: E. Ibar). 
Each target's exposure time {was} about 40min, reaching {an} rms of 
about 0.5 mK {at 111 km/s channel width.} 
The measured precipitable water vapour during the observations was about 0.8 (in {a} range 
{between} 0.6 {and} 1.2).
The APEX/SEPIA Band-5 data {were} reduced with CLASS software Version 1.1.
For every target, we trim{ed} the edge (about 3\%) of the spectrum and subtract{ed} 
the baseline by using a first-order polynomial.
Fig. {2} shows the CO line observations: 16 galaxies have {S/N$>3$ and were} fitted by single 
Gaussian profile (the top 12 spectra in Fig. {2} have {S/N$>5$}). 
8 galaxies have no CO detection down to 5$\sigma$ in the APEX spectra.
In Fig. {2} we also show the SDSS postage image
of each target in the upper-left corner of the CO spectrum.

\begin{figure}
\centering
\includegraphics[width=0.5\textwidth]{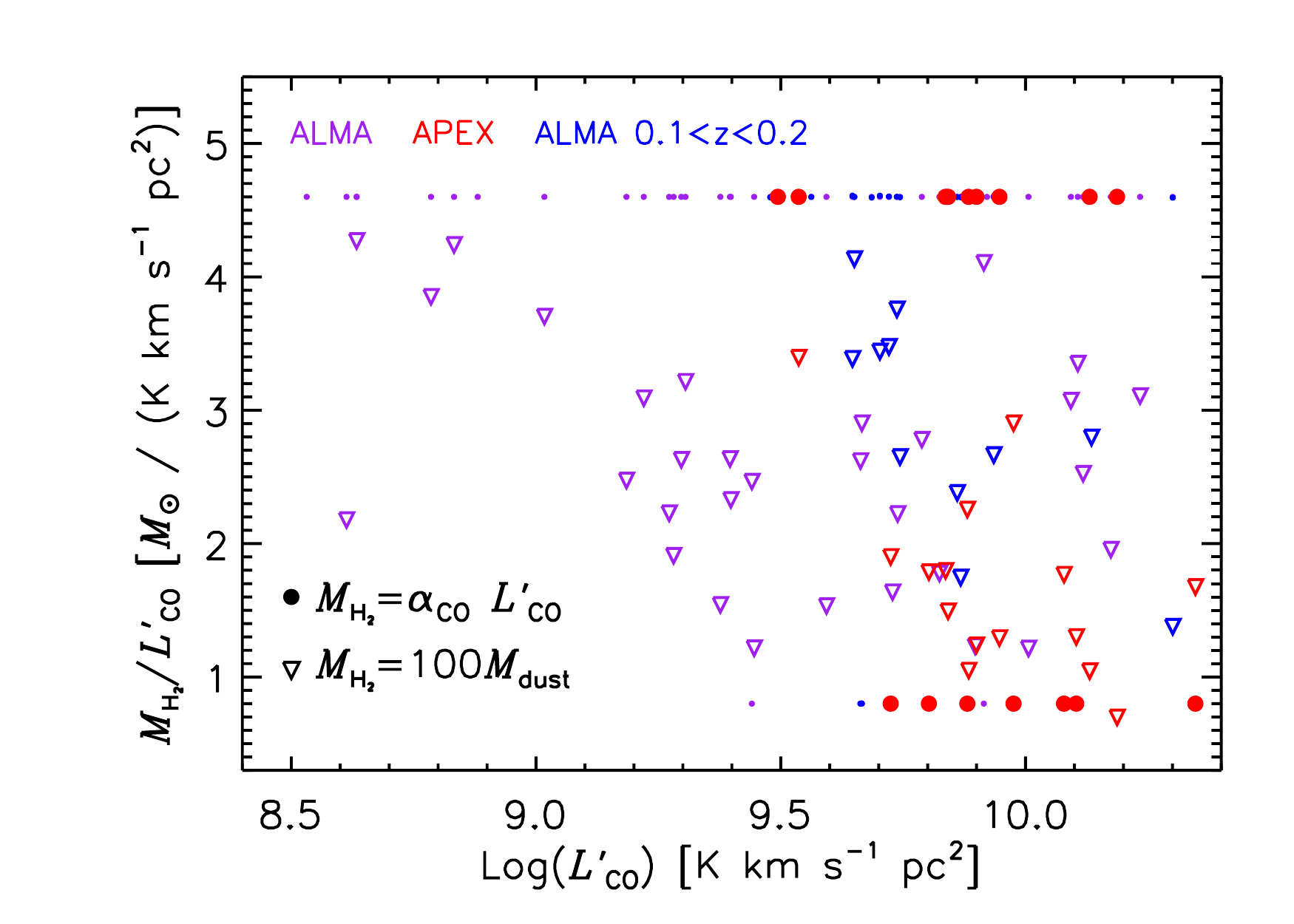}
\caption{The effective $\alpha_{\rm CO}$ estimate using different methods: the solid circles stand for the method of 
$M_{\rm H_2} = \alpha_{\rm CO} L'_{\rm CO}$ ($\alpha_{\rm CO} = 0.8\,M_{\odot}(\rm K\, km\, s^{-1} \,pc^{2})^{-1}$ 
for the mergers and $\alpha_{\rm CO} = 4.6\,M_{\odot}(\rm K\, km\, s^{-1} \,pc^{2})^{-1}$ for the isolated galaxies); 
the open triangles stand for the method of $M_{\rm H_2} = \delta M_{\rm dust}$ assuming the $\delta = 100$.
The ALMA data at 0.1$<$$z$$<$0.2 are represented by blue markers and those outside the range by purple ones.
}\label{ML}
\end{figure}
\begin{figure*}
\centering
\includegraphics[width=0.9\textwidth]{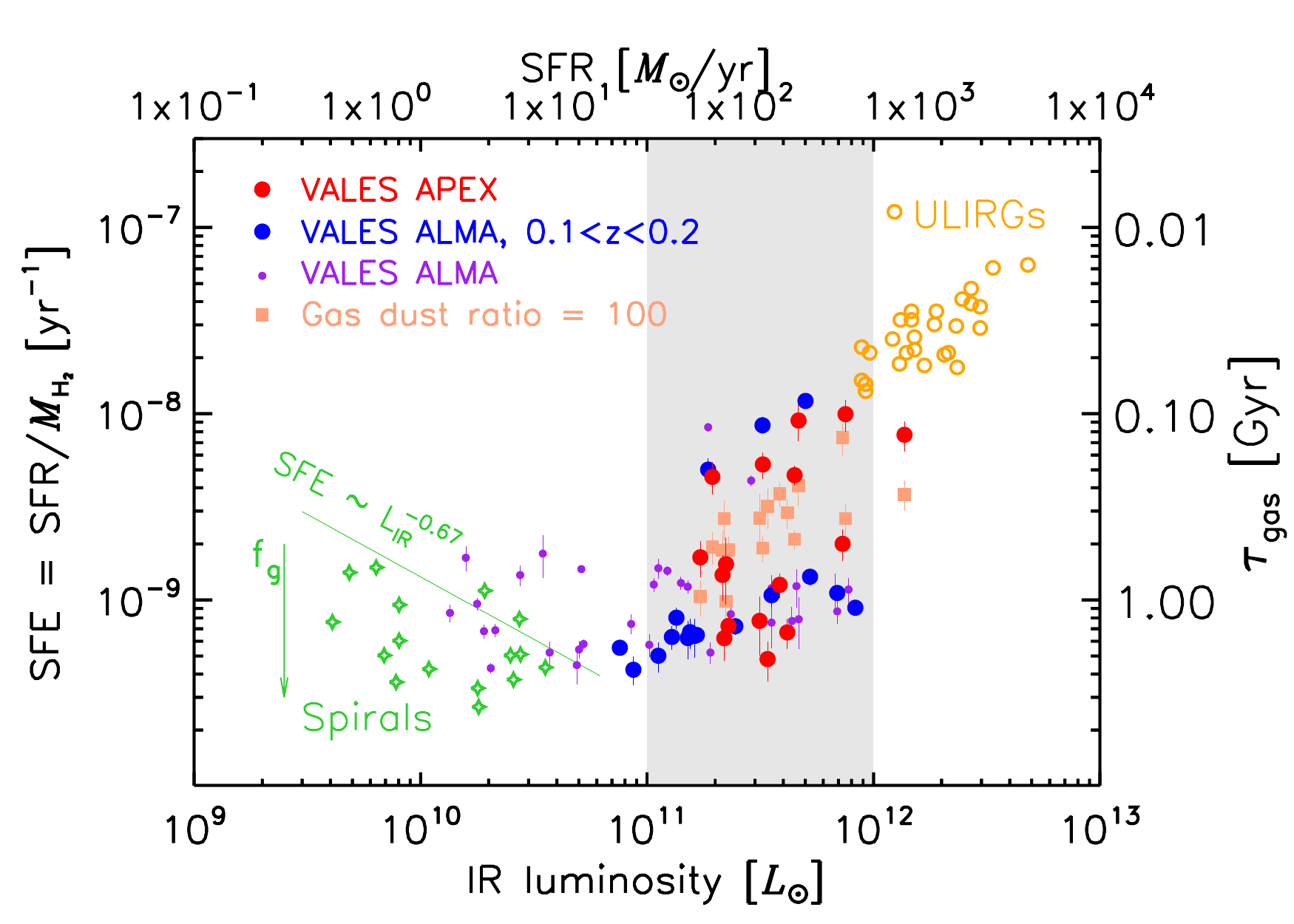}
\caption{The SFE vs $L_{\rm IR}$ relation. The colour and symbol codes are the same as Fig. {3}. 
We estimate the $M_{\rm H_2}$ with two methods: 
{assuming an $\alpha_{\rm CO}$ that depends on the optical morphology (solid circles)
and considering a simple gas to dust} mass ratio (solid squares). 
We adopt the $\alpha_{\rm CO} = 4.6 \,M_{\odot}(\rm K\, km\, s^{-1} \,pc^{2})^{-1}$ for the isolated galaxies 
and $\alpha_{\rm CO} = 0.8\,M_{\odot}(\rm K\, km\, s^{-1} \,pc^{2})^{-1}$ for the merging 
galaxies to derive the $M_{\rm H_2}$ from $L'_{\rm CO}$, as in previous papers of the VALES series.
The $H$-ATLAS data allow deriving the galaxy dust masses, which can be converted into H$_2$ mass 
by assuming the gas-dust ratio $\delta = M_{\rm gas}/M_{\rm dust} = 100$ (Magdis et al. 2012). 
Typical values for $\delta$ are in the range of 50 to 150, which can be affect the range of $M_{\rm H_2}$
by 0.3 dex. 
The SFE of the APEX data nicely bridge the gap between the normal star forming local galax{ies} and the local ULIRGs.
The shaded region corresponds to that where the SFE transitions occurs.
}\label{SFE}
\end{figure*}

\begin{figure}
\centering
\includegraphics[width=0.45\textwidth]{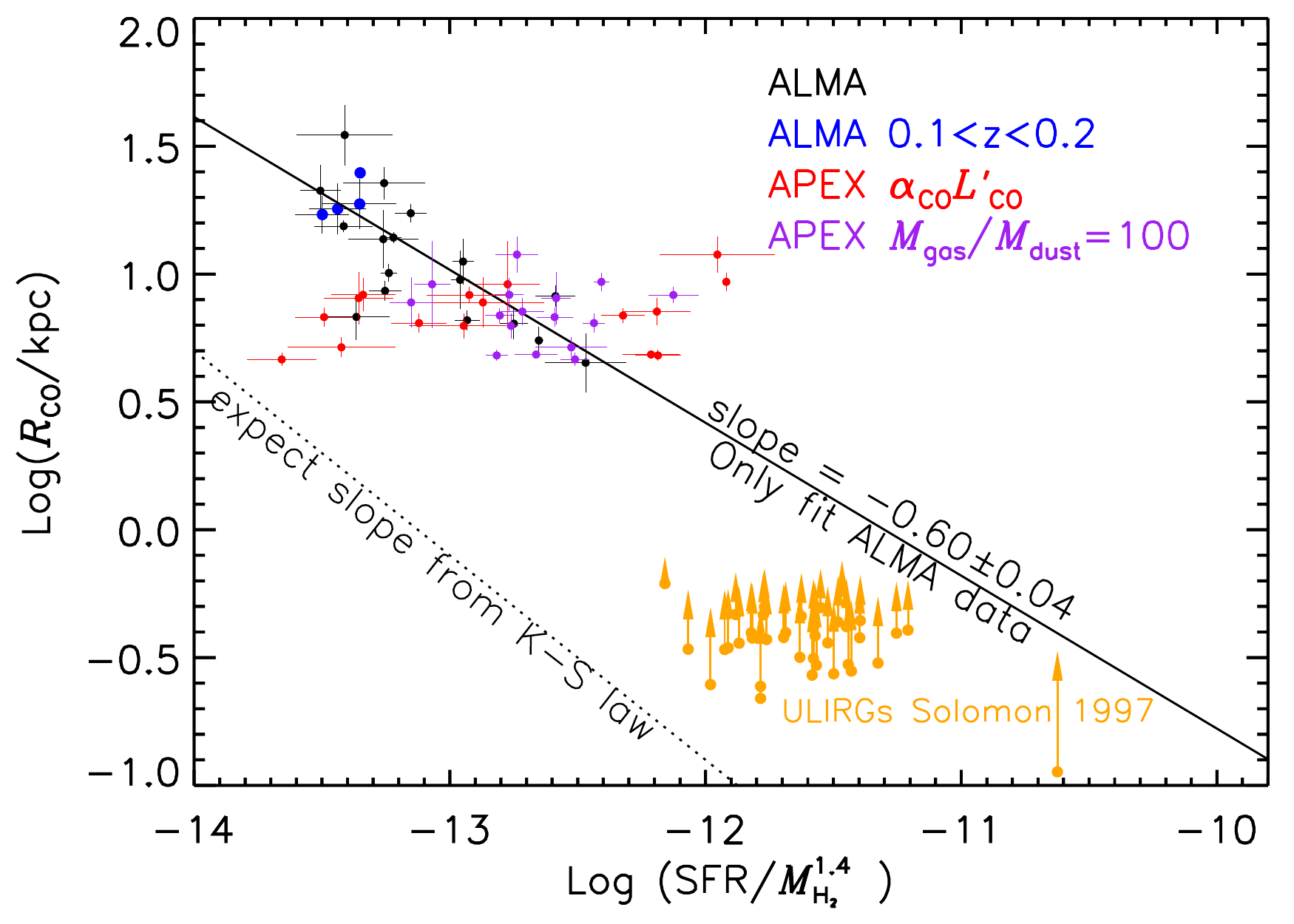}
\caption{.
The radius versus the ${\rm SFR} / M_{\rm gas}^{1.4}$ for all the ALMA resolved galaxies and 
the APEX targets. The ${\rm SFR} / M_{\rm gas}^{1.4}$ shows a monotonic relation with SFE. 
For the ALMA resolved galaxies, the slope of the correlation
is about -0.6$\pm$0.04. We also show the -0.8 slope {as a dashed} line, which is expected from the Kennicutt-Schmidt law. 
We do not have enough resolution for the APEX targets' CO radius ($R_{\rm CO}$), so we use their SDSS $r$ band Petrosian radii/1.6
instead (Villanueva et al. 2017). The orange dots show the lower limits of the radii
of ULIRGs sample of {the} Solomon et al. (1997).
}\label{R_CO}
\end{figure}

\begin{figure}
\includegraphics[width=0.5\textwidth]{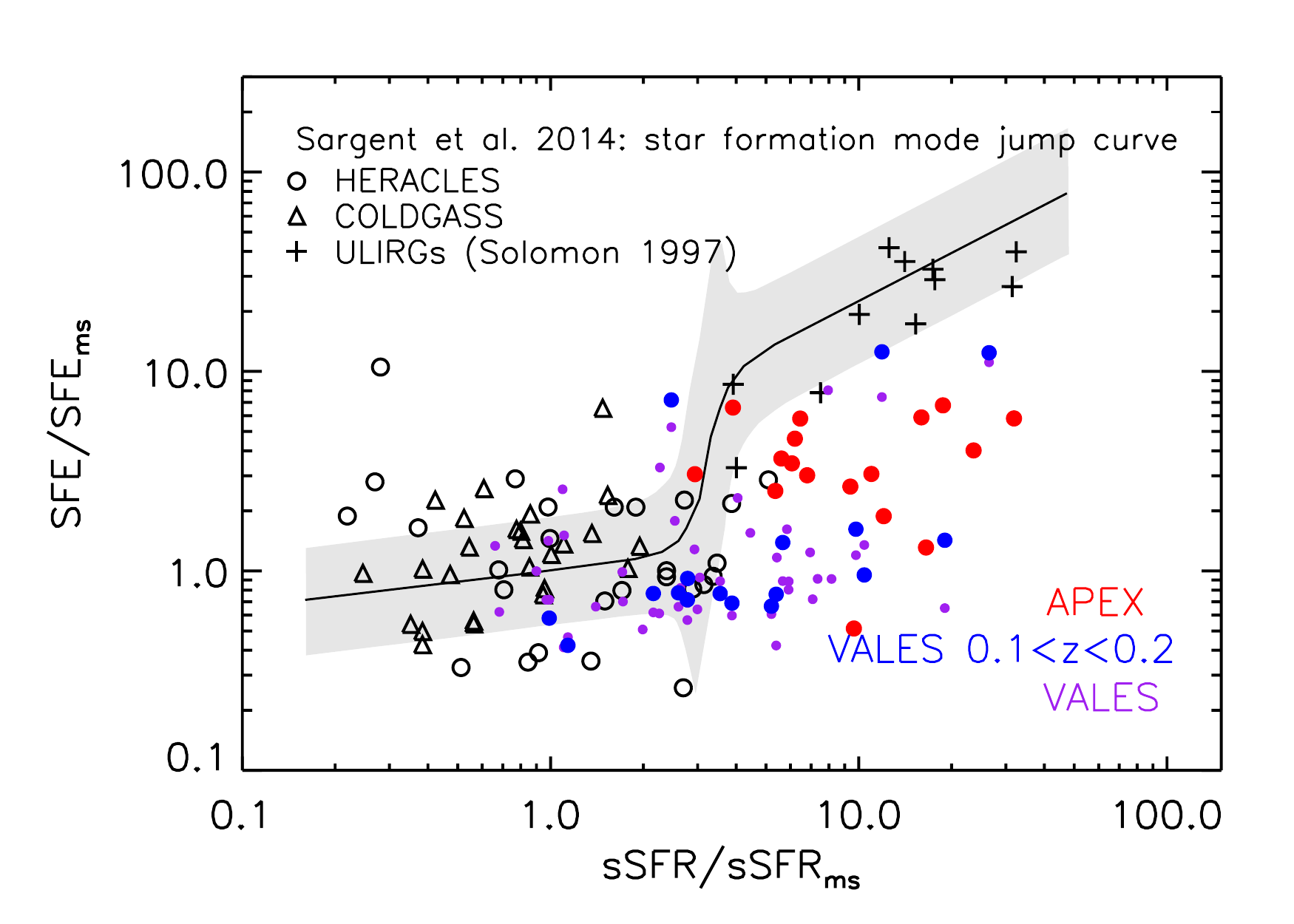}
\caption{{
The SFR-SFE relation as shown in Fig. 11 in Sargent et al. (2014). 
We employ the normal galaxies case Eq. 4 in Sargent et al. (2014) to derive the SFE/SFE$_{\rm ms}$
Here we only plot the representative samples in Sargent et al. (2014) and we show their result
by thick line as well as the uncertainty by the shaded region.
The VALES (blue and purple bullets) and APEX (red bullets) samples show a large range along
the sSFR/sSFR$_{\rm ms}$ and they are not only consistent with the local normal SFGs at the low sSFR/sSFR$_{\rm ms}$
end but also transition smoothly between the normal SFGs and the ULIRGs.}
}\label{SFRSFE}
\end{figure}

\subsection{Galaxy Sample over $0.1 < z < 0.2$ for this study}

We have combined these new SEPIA Band-5 observations to the VALES sample.
We select the 16 galaxies with APEX/SEPIA Band-5 detections together
with the 17 galaxies previously detected by ALMA and shown by Villanueva et al (2017) in the
same redshift range. These 33 galaxies between $0.1<z<0.2$ lie in the 
grey shaded region of Fig. {1}. The cosmological timescale within this redshift
range $\Delta z(0.1-0.2)$ is about 1 Gyr, 
which is about the typical gas depletion timescale of normal SFGs.
Thus the cosmological evolution of the galaxies in this sample can be neglected
for normal galaxies, although {many} starburst cycles could be expected for more
active galaxies. 

Based on the visual inspection {of SDSS images
of} the 16 APEX galaxies, we identify 7 {possibly interacting (mergers)} 
and 9 isolated galaxies. On the other hand, the 17 ALMA CO-detected galaxies include 3 mergers and 14 isolated galaxies. 
The stellar masses are derived using MAGPHYS (da Cunha et al. 2008) while the 
${\rm SFR}(M_{\odot}/{\rm yr}) = 10^{-10}L_{\rm IR}/L_{\odot}$ (Villanueva et al. 2017)  using the 
{$L_{\rm IR}(8-1000\mu$m)} derived from 
fitting the broad-band far-IR photometry (assuming a Chabrier IMF, Chabrier2003).
We follow the same procedure to derive galaxy properties as Villanueva et al. (2017). 
We list the main properties of the APEX observed targets in Table {1}. 
For the targets with no CO detection, we list the upper limit (5$\sigma$) of
the CO flux.

\section{Results}

We define ${\rm SFE}'={\rm SFR}/L'_{\rm CO}$ and present the observed 
${\rm SFE}'$ vs $L_{\rm IR}$ in Fig. {3}.
The $L'_{\rm CO} = L'_{{\rm CO}\, (J=1\text{--}0)}$ is defined as (Solomon \& Vanden Bout 2005):
\begin{equation}
	L'_{{\rm CO}\, (J=1\text{--}0)} = 3.25\times 10^{7}S_{\rm CO}\Delta v \, \nu_{\rm obs}^{-2}D_{\rm L}^{2}(1+z)^{-3}[\rm K \,km\, s^{-1}\,pc^{2}],
\end{equation}
where $S_{\rm CO}\Delta v$ is the velocity-integrated flux density in units of $\rm Jy\, km\, s^{-1}$, 
$\nu_{\rm obs}$ is the observed frequency of the emission line in GHz, $D_{\rm L}$ is the luminosity distance in 
Mpc and $z$ is the redshift.
We assume a $L_{{\rm CO}\, (J=2\text{--}1)}/L_{{\rm CO}\, (J=1\text{--}0)}$ ratio as 0.85, 
although we reckon this may vary from 0.5 to 1 for different source cases (Carilli and Walter 2013). 
We expect this systematic variation will affect our result by 0.3 dex at most.

For the ``normal'' galaxies at $2\times 10^9 < L_{\rm IR} < 10^{11}L_{\odot}$, the ${\rm SFE}'$ is roughly constant with 
scatter of about 0.5 dex, so the SFR and the cold molecular gas mass are roughly proportional to each other (Villanueva et al. 2017).
The constant $\rm SFE'$ indicates that the gas of the normal spiral galaxies is just enough to maintain a long depletion 
time scale. 

We fit the scaling relation between ${\rm SFE}'$ and IR luminosity
for all the $L_{\rm IR}>10^{11}L_{\odot}$ galaxies from the VALES {including our
new SEPIA band-5 detections} and local ULIRGs {taken from Solomon et al. (1997)}:
\begin{equation}	
	\frac{\rm SFE'}{10^{-9}\rm yr^{-1}} = (2.8\pm0.9) \times \left( \frac{L_{\rm IR}}{10^{11}L_{\odot}} \right)^{0.71\pm0.01}.
\end{equation}\label{SFEp}
The quantities $\rm SFE'$ and $L_{\rm IR}$ in Eq. {2} are not independent from each other, 
nevertheless, this scaling relation can help us to 
estimate the CO luminosity of low-$z$ {LIRGs and ULIRGs}.

Previous studies (Magdis et al. 2014; Sargent et al. 2014) have shown that the $\rm SFE'$ of the spirals 
and ULIRGs sample in Fig. {3} are {consistent} with the CO emission surveys at $z<0.3$, 
e.g., COLDGASS, EGNoG and PHIBSS. 
{So we restrict the sample shown in Fig. 3 to spirals and ULIRGs for clarity.}
The $\rm SFE'$ in Fig. {3} changes smoothly with the IR luminosity. 
We do not see a clear separation of SFR modes (Daddi et al. 2010). 
We note that our LIRG sample bridges the parameter space between the normal SFGs and the local powerful ULIRGs.

\section{Discussion}
\subsection{The origin of the SFE scatter}
To understand the true SFE distribution of our sample, we need to tackle the molecular gas content. 
We use the $M_{\rm H_2} = \alpha_{\rm CO} L'_{\rm CO}$ relation as the primary estimator. 
{To maintain consistency with previous VALES papers in this series, we} 
follow the choice of $\alpha_{\rm CO}$ based on the
optical SDSS morphology as in Villanueva et al. (2007): $\alpha = 0.8 \,M_{\odot}(\rm K\, km\, s^{-1} \,pc^{2})^{-1}$ 
for merger systems and $\alpha = 4.6 \,M_{\odot}(\rm K\, km\, s^{-1} \,pc^{2})^{-1}$ for 
isolated {disky or bulgy} galaxies (Solomon \& Vanden Bout 2005; Bolatto et al. 2013).

The H$_2$ mass can also be estimated from the gas-to-dust ratio. We assume a constant gas-to-dust ratio
$\delta = M_{\rm gas}/M_{\rm dust} = 100$ (Magdis et al. 2012) where $M_{\rm dust}$ is derived 
from the MAGPHYS fits. {The} typical range of the $M_{\rm dust} \sim 10^{7.5-8.6}M_{\odot}$ and the 
results are listed in Table 1. The results are illustrated in Fig. {4}. 
Observationally, {a} typical range of the $\delta$ is from about 50 to 150 (Magdis et al. 2012), 
hence this can change the $M_{\rm H_2}/L'_{\rm CO}$ from 
0.5 to 1.5 times of the current value. As an independent approach of estimating the molecular mass, 
the gas-to-dust ratio method shows a good agreement with the $\alpha_{\rm CO}L'_{\rm CO}$ estimations (see also
Hughes et al. 2017b).

The SFE results are illustrated in Fig. {5}. In this figure, we identify two populations of SFGs: 
one as an extension of the ``main sequence" galaxies
with higher SFRs up to $L_{\rm IR} \sim 10^{11.5}L_{\odot}$; {the other population has SFE} 
between the ``main sequence'' and ULIRGs with a clear trend towards high SFE at higher $L_{\rm IR}$. 
The SFE based on the gas-to-dust ratio method suggests a population {filling} the parameter space between
the normal spirals and ULIRGs (Fig. {5}). Our observations reveal 
the existence of a wide range of SFE (specially between $L_{\rm IR} \simeq 10^{11-12}L_{\odot}$), 
combining both star-formation modes: ``disk-like'' and ``starburst'' features.

Looking at the scatter of the correlation in Fig. {5}, we {find} that the SFE
should be anti-correlated with the galaxy gas radius\footnote{The Schmidt-Kennicutt law follows 
$\Sigma_{\rm SFR} \propto \Sigma_{\rm gas}^{1.4}$. If we make the rough assumption that the 
${\rm SFR} \simeq \Sigma_{\rm SFR}R^{2}$ and $M_{\rm gas}\simeq\Sigma_{\rm gas}R^{2}$, then the 
$\Sigma_{\rm SFR}R^{2} \propto \Sigma_{\rm gas}^{1.4} R^{2.8} R^{-0.8}$. So 
${\rm SFR} / M_{\rm gas}^{1.4} \propto R^{-0.8}$. Since the ${\rm SFR} / M_{\rm gas}^{1.4}$ is
monotonic with SFE, if the Kennicutt-Schmidt law is
valid for all the galaxies, galaxies with small gas radii would have higher SFE.}. Fig. {6} 
shows the ALMA resolved CO radius ($R_{\rm CO}$) against the ${\rm SFR} / M_{\rm gas}^{1.4}$, which 
highlights the importance of the $R_{\rm CO}$ in controlling the SFE.
For the more normal SFGs (resolved by ALMA at 3.5''), we find that the physical size{s} of the CO 
emitting regions tend to be larger for lower SFE, while compact CO emtters tend to have higher SFE. 
For all of the LIRGs observed by APEX (unresolved in CO),
we use the SDSS $r$ band Petrosian radius to estimate the $R_{\rm CO} \simeq R_{\rm r}/1.6$ following Villanueva et al (2017)
and show the results in Fig. {6}. We find that all of the APEX targets have similar {CO} radii, 
which is mainly caused by the narrow redshift range and the limited spatial resolution of SDSS.

For normal SFGs, the main sequence shows ${\rm SFR} \sim M_{*}^{\beta}$, 
where $\beta \simeq 0.6$ for local galaxies (Brinchmann et al. 2004).
The typical gas fractions ($f_{\rm g} = M_{\rm H_2}/(M_{*}+M_{\rm H_2})$) of local galaxy
samples are of about 10\%  (Narayanan et al. 2012b). 
If the SFR $\propto L_{\rm IR}$, then the combination of these three relations
will result in $SFE \propto (1/f_{\rm g}-1)L_{\rm _{IR}}^{\frac{\beta - 1}{\beta}}$. 
The $\beta$ index describes the slope of the SFE and 
$L_{\rm IR}$ relation while the $f_{\rm g}$ accounts for the {Y-axis intercept} (the green lines in Fig. {5}). 

Finally, considering that APEX measured the total CO flux within a primary beam of $\sim$35'' at $\sim$ 200GHz, 
which is larger than the typical size of a $z \sim 0.1$ galaxy, {in the case of merging systems}, 
close counterparts may also contribute to the CO flux. 
{This may be the case, for example, of HATLAS J121005.9+002639 where we see a possible broadening of the CO line 
emission. This broadening could be due to a close counterpart which is at the same redshift, as seen in Fig. 2.}
We also note that parameters such as $M_{*}$ could be affected too. Nevertheless, 
we {estimate} that the total stellar mass 
within the 35'' is {no more} than $\sim 2$ times larger than the stellar mass of the {central} galaxy. 
Thus the stellar mass will not significantly affect the main trends we find in our data.
\begin{table*}
	\centering
	\caption{Parameters of the APEX Observed CO (2-1) detected targets.}
	\label{tab}
	\begin{tabular}{lccccccccc} 
		\hline
		HATLAS ID & GAMAID & $z_{\rm spec}$ &$\log(M_{*}/M_{\odot})$&  $\log (L' _{\rm CO}[\rm K km\,s^{-1}pc^{2}])$ & $\log(L_{\rm IR}/L_{\odot})$ & $\log(M_{\rm dust}/M_{\odot})$ & $\rm CO\, FWHM$ km/s\\
		\hline
		HATLAS J114039.1-001821 & 53812   & 0.1128 &10.50 $\pm$ 0.03 &   9.70 $\pm$ 0.09 & 11.29 $\pm$ 0.01 & 8.00 $\pm$ 0.06 & 304$\pm$  59\\
		HATLAS J115447.6+012825 & 219701  & 0.1497 &10.57 $\pm$ 0.02 &   9.74 $\pm$ 0.12 & 11.86 $\pm$ 0.02 & 7.99 $\pm$ 0.04 & 233$\pm$  52\\
		HATLAS J115849.9-013146 & 185275  & 0.1468 &10.37 $\pm$ 0.03 &  10.13 $\pm$ 0.10 & 11.54 $\pm$ 0.03 & 8.03 $\pm$ 0.03 &1102$\pm$ 309\\
		HATLAS J120541.4-001420 & 55305   & 0.1763 &10.20 $\pm$ 0.03 &   9.74 $\pm$ 0.16 & 11.88 $\pm$ 0.03 & 8.44 $\pm$ 0.03 & 301$\pm$  60\\
		HATLAS J121005.9+002639 & 85450   & 0.1280 &10.93 $\pm$ 0.03 &  10.04 $\pm$ 0.06 & 11.65 $\pm$ 0.03 & 8.32 $\pm$ 0.05 & 535$\pm$  81\\
		HATLAS J141727.5-002535 & 568216  & 0.1226 &10.86 $\pm$ 0.11 &   9.93 $\pm$ 0.06 & 11.51 $\pm$ 0.02 & 8.23 $\pm$ 0.04 & 354$\pm$  54\\
		HATLAS J142225.2+002649 & 92214   & 0.1130 &10.53 $\pm$ 0.03 &   9.49 $\pm$ 0.15 & 11.33 $\pm$ 0.03 & 8.07 $\pm$ 0.06 & $<$185$\pm$  66\\
		HATLAS J142727.3-005842 & 544759  & 0.1623 &10.80 $\pm$ 0.03 &  10.35 $\pm$ 0.08 & 12.14 $\pm$ 0.05 & 8.57 $\pm$ 0.02 & 375$\pm$  69\\
		HATLAS J142831.9-003636 & 568985  & 0.1037 &9.80  $\pm$ 0.12 &  10.07 $\pm$ 0.09 & 11.23 $\pm$ 0.02 & 8.24 $\pm$ 0.06 & 526$\pm$ 137\\
		HATLAS J142948.7+010822 & 228482  & 0.1601 &10.35 $\pm$ 0.05 &   9.95 $\pm$ 0.15 & 11.50 $\pm$ 0.01 & 8.06 $\pm$ 0.06 & 504$\pm$ 167\\
		HATLAS J143155.0-005701 & 545019  & 0.1217 &11.05 $\pm$ 0.02 &   9.91 $\pm$ 0.08 & 11.62 $\pm$ 0.03 & 8.15 $\pm$ 0.05 & 450$\pm$ 120\\
		HATLAS J143334.2-012559 & 492771  & 0.1600 &10.37 $\pm$ 0.03 &   9.80 $\pm$ 0.10 & 11.67 $\pm$ 0.02 & 8.05 $\pm$ 0.04 & 347$\pm$  88\\
		HATLAS J143953.4+000618 & 79073   & 0.1321 &10.84 $\pm$ 0.06 &   9.49 $\pm$ 0.17 & 11.35 $\pm$ 0.01 & 8.36 $\pm$ 0.07 & 275$\pm$ 112\\
		HATLAS J144331.1-001624 & 64970   & 0.1417 &10.25 $\pm$ 0.07 &   9.88 $\pm$ 0.11 & 11.34 $\pm$ 0.01 & 7.90 $\pm$ 0.04 & 810$\pm$ 223\\
		HATLAS J144749.4+020209 & 343741  & 0.1193 &11.05 $\pm$ 0.10 &   9.76 $\pm$ 0.09 & 11.36 $\pm$ 0.01 & 8.09 $\pm$ 0.10 & 329$\pm$  62\\
		HATLAS J145008.3+015159 & 252158  & 0.1134 &10.92 $\pm$ 0.02 &   9.79 $\pm$ 0.08 & 11.59 $\pm$ 0.02 & 8.02 $\pm$ 0.03 & 355$\pm$  60\\
		HATLAS J120827.2+005704 &  99268  & 0.1591 &11.09 $\pm$ 0.02 &   $<$7.60         & 11.73 $\pm$ 0.01 &  8.30 $\pm$ 0.03  & $--$  \\
		HATLAS J142253.3+014355 & 319750  & 0.1104 &10.33 $\pm$ 0.03 &   $<$7.23         & 11.24 $\pm$ 0.05 &  7.83 $\pm$ 0.04  & $--$  \\
		HATLAS J113842.6-023316 & 123041  & 0.1050 &10.73 $\pm$ 0.09 &   $<$7.28         & 11.19 $\pm$ 0.04 &  7.85 $\pm$ 0.05  & $--$  \\
		HATLAS J121258.7-012123 & 145195  & 0.1042 &10.67 $\pm$ 0.02 &   $<$7.14         & 11.18 $\pm$ 0.02 &  7.89 $\pm$ 0.04  & $--$  \\
		HATLAS J122104.3+000506 &  71574  & 0.1071 &9.915 $\pm$ 0.03 &   $<$7.26         & 11.19 $\pm$ 0.04 &  7.75 $\pm$ 0.05  & $--$  \\
		HATLAS J114244.3-005450 & 534898  & 0.1076 &9.940 $\pm$ 0.03 &   $<$7.35         & 11.03 $\pm$ 0.06 &  7.79 $\pm$ 0.06  & $--$  \\
		HATLAS J142128.2+014845 & 319694  & 0.1604 &10.34 $\pm$ 0.04 &   $<$7.61         & 11.21 $\pm$ 0.02 &  8.41 $\pm$ 0.11  & $--$  \\
		HATLAS J121623.4+010614 & 24056   & 0.1552 &9.298 $\pm$ 0.10 &   $<$7.40         & 10.72 $\pm$ 0.03 &  8.49 $\pm$ 0.10   & $--$  \\
		\hline
	\end{tabular}
\end{table*}

\subsection{Confronting the bimodality in star formation efficiencies}

{Previous studies have shown that the SFE in SFGs follow a bimodal
behaviour for normal and starburst galaxies (Daddi et al. 2010; Genzel et al. 2010). 
Based on the existence of the `main sequence' followed
by normal SFGs, Sargent et al. (2014) show that this bimodal
SFE can be nicely illustrated by plotting SFE/SFE$_{\rm ms}$ versus
sSFR/sSFR$_{\rm ms}$, where SFE/SFE$_{\rm ms}$ is the SFE normalised
by the SFE of the main sequence or strongly starbursting galaxies with the same mass.
In Fig. {7} we show our VALES
(including the new APEX/SEPIA Band-5 detections) sample in comparison
with other samples taken from the literature, overplotted over the
predictions (solid line) by Sargent et al. (2014).  We find
that the steep jump in SFE at sSFR/sSFR$_{\rm ms}$\,$>$\,3 is not as
clean as expected. We highlight that our sample covers a much wider
range of sSFRs, with IR luminosities in the LIRGs range in between of
those normal (HERACLES and COLDGASS) and local ULIRGs. Indeed, it is
evident that Sargent et al. (2014)'s predictions are
significantly affected by low number statistics of the most strongly 
starbursting galaxies (specially at sSFR/sSFR$_{\rm ms}$\,$>$\,3). Our study
clearly shows that the most `starbursty' galaxies (see VALES range in
Fig. {1}) present a wide range of SFEs, i.e.\ not all galaxies
with high sSFR are passing through a dominant starburst phase. As
shown in Fig. {3}, there is a smooth transition of SFEs as
a function of IR luminosities, in contradiction with a bimodal
behaviour.}

{Our evidende is also supported by a recent study by 
Lu et al. (2017), using local LIRGs, that shows a similarly smooth
transition in SFEs as analysed by the [CI]370$\mu$m/CO(7-6) and
C(60$\mu$m/100$\mu$m) ratios (Fig. 17, panel c in their paper). 
Here we will consider the CO(7-6) as a
proxy for the SFR and [CI]370$\mu$m for the $M_{\rm H_{2}}$
(Jiao et al. 2017). {Lu et al. (2017)} suggest that the change in
effective SFE could be correlated to the far-IR colour, i.e. the dust
temperature. On the other hand, a smooth transition of the SFE is also
found by the recent study by Lee et al. (2017) using a sample
of 20 intermediate redshift ($0.25<z<0.65$) LIRGs in the 
Cosmological Evolution Survey (COSMOS).
We conclude that our findings are
consistent with previous studies of LIRGs at low- and
intermediate-redshifts.}

{We have identified different parameters responsible for the observed
scatter of the SFE in SFGs, nevertheless the only parameter which most
probably causes the appearance of a bimodal behaviour comes from the
bimodal assumption of the $\alpha_{\rm CO}$ conversion factor. It is
well known that $\alpha_{\rm CO}$ is dependent on the physical
properties of the ISM (e.g.\ metallicity, density, temperature; Solomon \& Vanden Bout 2005), 
so that making a global assumption for a single $\alpha_{\rm CO}$ value over the whole galaxy
can be significantly affected by complex systematical uncertainties (e.g. Sandstrom et al. 2013).}

\section{conclusion}

Our new APEX/SEPIA Band-5 observations double the number of sources with CO detections in the ALMA-based 
VALES sample at $0.1 < z < 0.2$, specially covering the parameter space at higher sSFR. 
In this work we concentrate in an investigation of the global star formation efficiency of 
CO-detected galaxies, including previous low-z surveys taken from the literature. 
To avoid the uncertainties on the $\alpha_{\rm CO}$ conversion factor, we explore the correlation between 
the ${\rm SFR}/L'_{\rm CO}$ ratio and the IR luminosity. 
We find that this ratio remains {relatively} constant up to $L_{\rm IR}\sim 10^{11}L_{\odot}$ (a scatter of $\sim$\,0.5 dex), 
although above this value there is a clear increment in the effective SFE. 
{Benefited by the large sSFR/sSFR$_{\rm ms}$ of our VALES and APEX sample, we find a 
smooth transition of the SFE instead of a steep jump from the normal SFGs to the ULIRGs.}
{The smooth increment as a function of far-IR luminosity (specially between $10^{11-12}L_{\odot}$) is 
consistent with the previous LIRGs study (Lu et al. 2017; Lee et al. 2017).}
This suggests that the dominating star-formation mechanism (starburst or disk-like) 
smoothly changes between powerful ULIRGs and normal galaxies. We conclude that the main parameters 
controlling the scatter of the global SFE versus LIR correlation are: 
the assumed $\alpha_{\rm CO}$ conversion factor, the gas fraction and the physical size of the galaxies.

\section*{Acknowledgements}
This paper benefited from a number of thoughtful comments made by the anonymous referee.
This work was support from the Chinese Academy of Sciences (CAS) through the CASSACA Postdoc Grant and 
the Visiting Scholarship Grant administered by the CAS South America Center for Astronomy (CASSACA), NAOC.
E.I. and T.M.H. acknowledge the CONICYT/ALMA funding Program in Astronomy/PCI Project N$^{\circ}$: 31140020.
E.I. acknowledges partial support from FONDECYT through grant N$^\circ$\,1171710.
T.M.H. acknowledges the support from
the Chinese Academy of Sciences (CAS) and the National Commission for
Scientific and Technological Research of Chile (CONICYT) through a
CAS-CONICYT Joint Postdoctoral Fellowship administered by the CAS
South America Center for Astronomy (CASSACA) in Santiago, Chile.
R.L. acknowledges the support from Comit\'e Mixto ESO-GOBIERNO DE CHILE and GEMINI-CONICYT FUND 32130024.
G.O. acknowledges the support provided by CONICYT (Chile) through FONDECYT postdoctoral research grant no 3170942.
C.K.X. acknowledges the support of NSFC-11643003.
N.L. acknowledges the support by the NSFC grant \#11673028 and by the National Key R\&D Program of China
grant \#2017YFA0402704. 
Y.Q.X. acknowledges the support of NSFC-11473026 and 11421303.
This publication is based on data acquired with the Atacama Pathfinder Experiment (APEX). APEX is a collaboration between the Max-Planck-Institut fur Radioastronomie, the European Southern Observatory, and the Onsala Space Observatory.
This paper makes use of the following ALMA data: ADS/JAO.ALMA 2012.1.01080.S \& ADS/JAO.ALMA 2013.1.00530.S. ALMA is a partnership of ESO (representing its member states), NSF (USA) and
NINS (Japan), together with NRC (Canada), NSC and
ASIAA (Taiwan), and KASI (Republic of Korea), in cooperation with the Republic of Chile. The Joint ALMA
Observatory is operated by ESO, AUI/NRAO and NAOJ.





\bsp	
\label{lastpage}
\end{document}